# Colour Guided Colour Image Steganography

R.Amirtharajan, Sandeep Kumar Behera, Motamarri Abhilash Swarup, Mohamed Ashfaaq K
and John Bosco Balaguru Rayappan

Department of Electronics & Communication Engineering
School of Electrical & Electronics Engineering
SASTRA University
Thanjavur, Tamil Nadu, India
amir@ece.sastra.edu

*Abstract*— **Information security has become a cause of concern because of the electronic eavesdropping. Capacity, robustness and invisibility are important parameters in information hiding and are quite difficult to achieve in a single algorithm. This paper proposes a novel steganography technique for digital color image which achieves the purported targets. The professed methodology employs a complete random scheme for pixel selection and embedding of data. Of the three colour channels (Red, Green, Blue) in a given colour image, the least two significant bits of any one of the channels of the color image is used to channelize the embedding capacity of the remaining two channels. We have devised three approaches to achieve various levels of our desired targets. In the first approach, Red is the default guide but it results in localization of MSE in the remaining two channels, which makes it slightly vulnerable. In the second approach, user gets the liberty to select the guiding channel (Red, Green or Blue) to guide the remaining two channels. It will increase the robustness and imperceptibility of the embedded image however the MSE factor will still remain as a drawback. The third approach improves the performance factor as a cyclic methodology is employed and the guiding channel is selected in a cyclic fashion. This ensures the uniform distribution of MSE, which gives better robustness and imperceptibility along with enhanced embedding capacity. The imperceptibility has been enhanced by suitably adapting optimal pixel adjustment process (OPAP) on the stego covers.**

## I. INTRODUCTION

Communication skills have always been the hallmark of human interactivity. The primitive techniques that include cave drawings, smoke signal, drums etc. ascertain that over the years has been the modus operandi (mode of operation) of social & commercial intercourse. The technological proliferation in this electronic epoch has expanded horizons and has empowered organizations, nations or corporations to share intellectual information and be mutually benefitted by it. But with the rats race to acquire power, morals have devolved and has made electronic eavesdropping a prime problem. Any popular system of governance ranging from the banking sector to the administrative sector of a country must be behind the barrages of efficacious security systems to protect it from apathetic and amoral people who accomplice to pull out any valuable information they can grab. Data encryption [1] and Data hiding [2, 3] techniques are potential tools for securing sensitive information and hence is widely used to protect the data over an overt channel from malicious attackers.

In data encryption the sender encodes the data using a key K. Only the legitimate receiver will have the key K to decrypt or decipher the code to get back the original message. But as the encoded data is in the form of a string of abstract codes, it will intrigue hackers. Steganography is a discipline of data encryption and has over the years surmounted the problem.

In today's info-driven world, with accrooment of confidential information, there has been a corresponding increase in the attempts to sabotage the security guards of such information. This has led to an avalanche of pro and anti security innovations. Steganography [2] is one such pro-security innovation in which secret data is embedded in a cover. The concept of data hiding or steganography was first introduced by Simmons in 1983 [4]. Dictionary defines "Steganography" as "the art of writing in cipher, or in character, which are not perceivable except to person who has the key". In the field of computers, steganography has evolved as the promising option of hiding a message in a cover, whose presence cannot be discerned by any third party without the knowledge of the key. The cover in which the message is hidden can either be a text, image, audio or video file. Even after hiding the data, the stego image should be imperceptible i.e., the cover image and the stego image should be inert and impregnable.

Steganography has inherent strengths such as detection of minuscule and immense integrity, which can be further strengthened by introducing controlled variations in the pattern of embedding of secret data. When these controlled variations are in the form of difference in encoding of the secret data and a pseudo-random choice of traversal path [5, 6], the detection becomes very much lesser than that which can be obtained using many established Steganographic algorithms. The encryption of secret data prior to embedding increases the security to a great extent [2].

The rest of the paper is organized as follows. Section II describes the related works. Section III, discuss in detail about the three proposed scheme for embedding the secret data into a cover image. Section IV explains the simulation results for the proposed algorithm for various images. Section V provides the conclusion.

## II. RELATED WORKS

History provides us with a plethora of data hiding methods [2, 3, 7, 8, 9]. The first manifestation of steganography dates back to the Greeks. Herodotus passed information on wax covered tablets. Pirate legends tell of the practice of tattooing secret information, such as a map, on the head of someone, so that the hair would conceal it. Another common form is the use of invisible ink derived from vinegar, fruit juices and milk. In World War II, hidden sensors, microdot and grill methods were widely used. During civil war, there was a method of providing secret messages to slaves to aid their escape through quilt patterns. Presently watermarking [2] and spread spectrum techniques [2, 10] are gaining momentum.

### A. Classification of information system:

The generic classification of steganography [11]is : Based on the domain: first is spatial domain [2-9, 14, 15 ], second is transform domain [2, 12, 13, 14]. The former employs the hiding of the secret data in the pixels of the cover image and the latter employs the hiding of the data in the transform domain of the host. In the spatial domain the secret information is hidden in the pixels of cover image by employing Least Significant Bit (LSB) [2-9,14,15 ], pixel value differencing [16,17] and mod [9,16]. The transform functions the discrete cosine transform (DCT) [12] and discrete wavelet transform (DWT) [13] are employed to convert the pixel values into transform domain. Then the embedding of the data is done. Another classification is based on the exchange of the stego key [2] and it's a borrowed concept from cryptography [1]:

Pure steganography does not necessitate the exchange of a stego key as it is based upon the presumption that no other party is aware of this secret message. In secret key steganography, the stego key has to be exchanged prior to communication. In public key steganography a public key and private key are used to secure the communication between the parties who desire to communicate secretly.

There are several other approaches in classifying steganographic systems [2, 3, 7, 8, 11] based on the type of covers or according to the cover modifications applied in the embedding process.

The aforementioned classification based on the cover [2, 3] is Audio steganography, video steganography, Image steganography and text steganography. The other classification is based on the methodology or techniques as follows:

- Substitution systems substitute redundant parts of the cover with a secret message

- Transform domain techniques embed secret information in a transform space of the signal (e.g., in the frequency domain)

- Spread Spectrum techniques adopt ideas from spread spectrum communication

- Statistical Methods encode information by changing several statistical properties of a cover and use hypothesis testing in the extraction process

- Distortion techniques store information by signal distortion and measure the deviation from the original cover in the decoding step

- Cover generation methods encode information in a way that the cover for the secret communication is created.

Yet another classification of information hiding [10] called digital watermarks, also known as fingerprinting is significant especially in copyrighting materials. The watermarks are overlaid in files, which appear to be part of the original file and are thus not easily detectable by any average person. It has been widely used to protect the copyright of digital images as it embeds a trademark of the owner into the protected image, thus allowing the owner to prove their ownership of the suspected image by retrieving the embedded trademark. Water marking must satisfy the requirements of imperceptibility, security and robustness.

## III. THE PROPOSED METHODOLOGY

The proposed methodology uses the same principle of least significant bit insertion (LSB) along with a modified version of the Pixel Indicator method [18, 19]. Each pixel value of a colour image is represented by three bytes i.e. to define the intensity of the channels RED, BLUE, GREEN. One of the channels (RED, BLUE, GREEN) is used to indicate, how many numbers of binary data has to be hidden in the remaining two channels. Where in [18] authors proposed a Pixel Indicator methodology but the number of bits in each pixel is simple K bits LSB based steganography, which is not adaptive. In the stated methodology [19], the number of bits to be embedded in each channel is decided by Pixel value differencing (PVD) [17] and is also going to be guided by the indicator channel similar to [18]. The randomization, improved imperceptibility and robustness of the system are also simultaneously achieved by hiding the secret data in LSBs of the pixels [9], with more randomization in selection of the number of bits to be hidden and the channels in which the data is to be hidden [18]. The problem in [19] is moderate embedding capacity but it improves the imperceptibility with increased randomness.

On the grounds of the above mentioned techniques we have formulated three methodologies which can be applied on to colour images. In the first methodology, RED is used as default pointer. Here, Excess 3 value of the last 2 bits of RED channel decides the number of secret data bits hidden in BLUE and GREEN channels. Paper [5] suggested that embedding more number of bits in blue than green will give better performance hence it is adapted and shown in Table I. In the second methodology, user gets to select any channel as pointer however embedding procedure remains the same as proposed in methodology one. The third methodology is to improve security and the image quality as the pointer channel is not fixed. Instead the pointer is chosen based on a cyclic sequence and the data is embedded as per the table 1. i.e. in the first pixel Red is the pointer, while Green is channel 1 and Blue is the channel 2, in the second pixel, Green is the pointer, while Red is channel 1 and Blue is channel 2 and in third pixel Blue is the pointer, while Red is channel 1 and Green is channel 2. After doing this we apply OPAP (Optimal pixel adjustment process) [9] which checks each channel of each embedded pixel and modify all the 'k+1' bit to reduce the MSE in the channel if the modification gives better results. where 'k' is the number of bits embedded in that channel of the pixel.

Table I : Indicator value and its meaning.

| INDICATOR | CHANNEL 1 | CHANNEL 2 |
|-----------|-----------|-----------|
| 00 | 1 bit of data embedded | 2 bits of data embedded |
| 01 | 2 bits of data embedded | 2 bits of data embedded |
| 10 | 2 bits of data embedded | 3 bits of data embedded |
| 11 | 3 bits of data embedded | 3 bits of data embedded |

### A. FLOWCHART For Embedding:

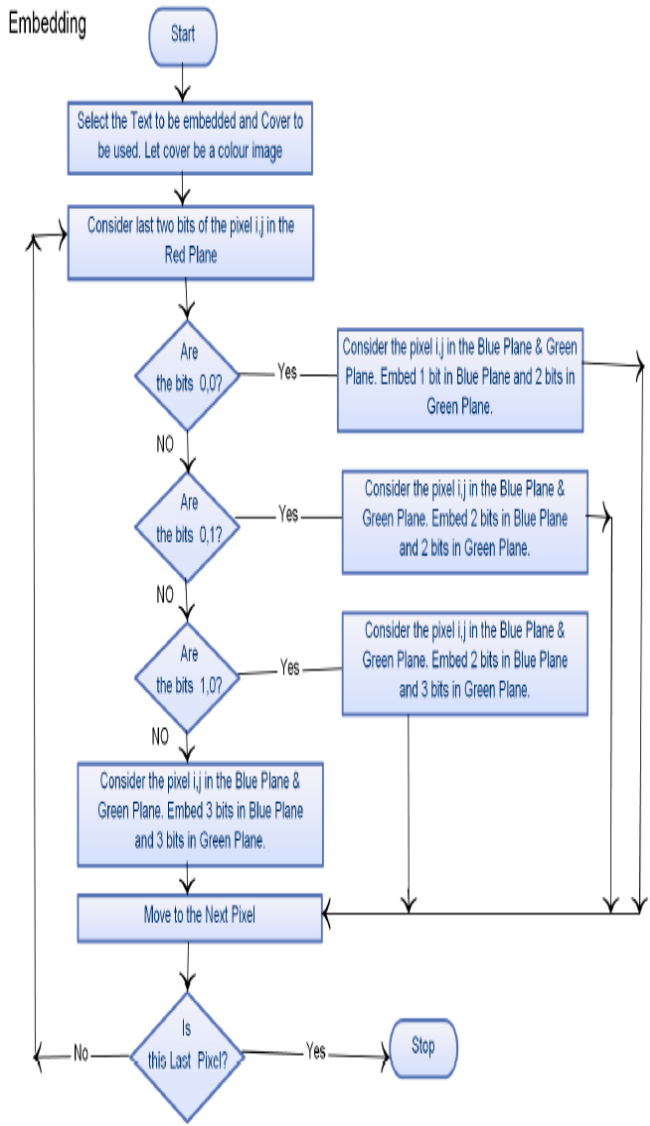

### B. FLOWCHART For Extraction:

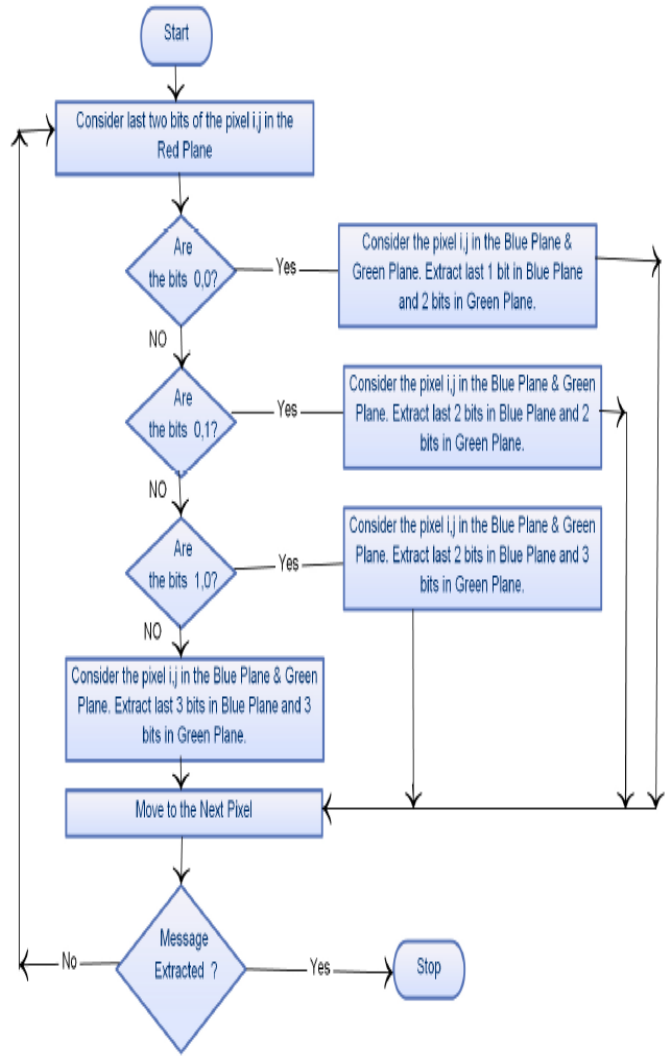

### C. METHOD -1 Algorithm:

**Embedding Algorithm:**

Inputs: Secret Data (D), Cover Image(C)
Output: Stego image(S) with secret data embedded in it.

1. Convert the Secret Data (D) into binary format.

2. Split the cover image C into Red, Green and Blue Planes.(R,G and B respectively)

3. For each pixel in R, do the following:
    3.1. Let b[0]=LSB of the current pixel in R
    3.2. Let b[1]=Next LSB of the current pixel in R
    3.3. Let n= (Decimal value of b) + 3

i.e., (Excess 3 value of b)
    3.4. If (n mod 2 = 0) then

Embed (n/2) bits of secret data in current pixels of G and B.

Else

Embed [(n-1)/2] bits and [(n+1)/2] bits of secret data in current pixels of G and B respectively.

3.5. If all secret data is embedded, then
Go to step-4

4. Store the resulting image as Stego Image (S) after applying OPAP.

**Recovery Algorithm:**

Input: Stego Image(S)
Output: Secret Data (D)

1. Split the stego image S into Red, Green and Blue Planes.(R,G and B respectively)

2. For each pixel in R, do the following:
2.1. Let b[0]=LSB of the current pixel in R
2.2. Let b[1]=Next LSB of the current pixel in R
2.3. Let n= (Decimal value of b) + 3

i.e., (Excess 3 value of b)
2.4. If (n mod 2 = 0) then
Read (n/2) LSBs of current pixels of G and B concatenate to D.

Else

Read [(n-1)/2] bits and [(n+1)/2] LSBs of current pixels of G and B respectively and concatenate to D.

3. Store the resulting recovered secret data (D).

*D. METHOD-2 Algorithm:*

**Embedding Algorithm:**

Inputs : Secret Data(D), Cover Image(C), Indicator-plane Index(I)
Output: Stego image(S) with secret data embedded in it.

1. Convert the Secret Data (D) into binary format.

2. Split the cover image C into Red, Green and Blue Planes.(R,G and B respectively)

3. If I=1 then,
P[1]=R, P[2]=G, P[3]=B
Else if I=2, then

P[1]=G, P[2]=R, P[3]=B
Else if I=3, then

P[1]=B, P[2]=R, P[3]=G

3. For each pixel in P[1], do the following:
3.1. Let b[0]=LSB of the current pixel in P[1]
3.2. Let b[1]=Next LSB of the current pixel in P[1]
3.3. Let n= (Decimal value of b) + 3

i.e., (Excess 3 value of b)
3.4. If (n mod 2 = 0)
Embed (n/2) bits of secret data in current pixels of P[2] and P[3].

Else

Embed [(n-1)/2] bits and [(n+1)/2] bits of secret data in current pixels of P[2] and P[3] respectively.
3.5. If all secret data is embedded, then
Go to step-4
4. Store the resulting image as Stego Image (S) after applying OPAP.

**Recovery Algorithm:**

Input : Stego Image(S), Indicator-plane index (I)
Output: Secret Data (D)

1. Split the stego image S into Red, Green and Blue Planes.(R,G and B respectively)
2. If I=1 then,
P[1]=R, P[2]=G, P[3]=B
Else if I=2, then

P[1]=G, P[2]=R, P[3]=B
Else if I=3, then

P[1]=B, P[2]=R, P[3]=G

3. For each pixel in P[1], do the following:
3.1. Let b[0]=LSB of the current pixel in P[1]
3.2. Let b[1]=Next LSB of the current pixel in P[1]
3.3. Let n= (Decimal value of b) + 3

i.e., (Excess 3 value of b)
3.4. If (n mod 2 = 0) then
Read (n/2) LSBs of current pixels of P[2] and P[3] and concatenate to D.

Else

Read [(n-1)/2] bits and [(n+1)/2] LSBs of current pixels of P[2] and P[3] respectively and concatenate to D.
4. Store the resulting recovered secret data (D).

*E. METHOD-3 Algorithm:*

**Embedding Algorithm:**

Inputs: Secret Data(D), Cover Image(C)
Output: Stego image(S) with secret data embedded in it.

1. Convert the Secret Data (D) into binary format.

2. Split the cover image C into Red, Green and Blue Planes.(R,G and B respectively)
3. Let index i=1.
4. For each pixel in P[1], do the following:
    4.1. If (i mod 3) =1 then,
        I[i]=1
        Else if (i mod 3)=2 then,
        I[i]=2
        Else
        I[i]=3
    4.2. Set i=i+1
5. Let index j=0
6. For each pixel in P[1], do the following:
    6.1. If I[j]=1 then,
        P[1]=R[i], P[2]=G[i], P[3]=B[i]
        Else if I[j]=2, then

P[1]=G[i], P[2]=R[i], P[3]=B[i]
        Else if I[j]=3, then

P[1]=B[i], P[2]=R[i], P[3]=G[i]

    6.2. Let b[0]=LSB of  P[1]
    6.3. Let b[1]=Next LSB of P[1]
    6.4. Let n= (Decimal value of b) + 3

i.e., (Excess 3 value of b)
    6.5. If (n mod 2 = 0) then
        Embed (n/2) bits of secret data in current pixels of P[2] and P[3].
        Else
        Embed [(n-1)/2] bits and [(n+1)/2] bits of secret data in current pixels of P[2] and P[3] respectively.
    6.5. If all secret data is embedded, then
        Go to step-7
        Else
        j = j+1
7. Store the resulting image as Stego Image (S) after applying OPAP.

**Recovery Algorithm:**

Input : Stego Image(S)
Output:Secret Data (D)

1. Split the stego image S into Red,Green and Blue Planes.(R,G and B respectively)
2. Let index i=1.
3. For each pixel in P[1], do the following:
    3.1. If (i mod 3) =1 then,
        I[i]=1
        Else if (i mod 3)=2 then,
        I[i]=2
        Else
        I[i]=3
    3.2. Set i=i+1

4. Let index j=0
5. For each pixel in P[1], do the following:
    5.1. If I[j]=1  then,
        P[1]=R[i], P[2]=G[i], P[3]=B[i]
        Else if I[j]=2, then

P[1]=G[i], P[2]=R[i], P[3]=B[i]
        Else if I[j]=3, then

P[1]=B[i], P[2]=R[i], P[3]=G[i]

    5.2. Let b[0]=LSB of  P[1]
    5.3. Let b[1]=Next LSB of P[1]
    5.4. Let n= (Decimal value of b) + 3 i.e., (Excess 3 value of b)
    5.5. If (n mod 2 = 0) then
        Read (n/2) LSBs of current pixels of P[2] and P[3] and concatenate to D.
        Else
        Read [(n-1)/2] bits and [(n+1)/2] LSBs of current pixels of P[2] and P[3] respectively and concatenate to D.
6. Store the resulting recovered secret data (D).

The effectiveness of the stego process proposed has been studied by estimating the following two metrics for the digital images.

    *F. Peak Signal to Noise Ratio (PSNR):*

The PSNR is calculated using the equation

$$PSNR = 10\log_{10}\left(\frac{I_{max}^2}{MSE}\right)dB \qquad (1)$$

where $I_{max}$ is the intensity value of each pixel which is equal to 255 for 8 bit gray scale images. Higher the value of PSNR better the image quality

    *G. Mean Square Error (MSE)*

The MSE is calculated by using the equation,

$$MSE = \frac{1}{MN}\sum_{i=1}^{M}\sum_{j=1}^{N}\left(X_{i,j}-Y_{i,j}\right)^2 \qquad (2)$$

IV.    EXPERIMENTAL RESULTS & DISCUSSION

In this present implementation Lena, baboon, Gandhi and Temple of 256 × 256 color digital images has been taken as cover images as shown in Figure 1 a, b, c & d and tested for full embedding capacity and the results are given in Figure 2, 3 and 4. The effectiveness of the stego process proposed has been studied by calculating MSE and PSNR for all the four digital images in RGB planes using the proposed methods I, II and III as given in Table II, III and IV

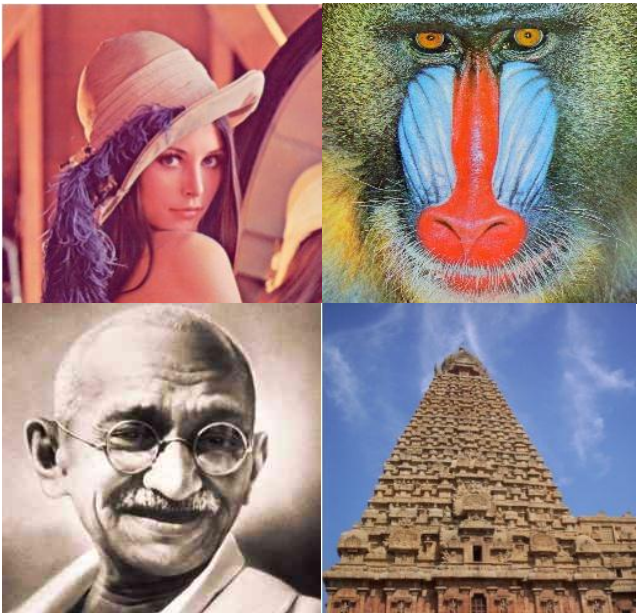

Figure 1 a. Lena b. Baboon c. Gandhi d. Temple

Table II: Method -1

| Cover Image | Channel I Red | | Channel II Green | | Channel III Blue | | BPP (Bits Per Pixel) |
|---|---|---|---|---|---|---|---|
| | MSE | PSNR | MSE | PSNR | MSE | PSNR | RGB |
| Lena | 0 | ∞ | 0.75 | 49.37 | 1.17 | 47.45 | 4.51 |
| Baboon | 0 | ∞ | 0.76 | 49.32 | 1.17 | 47.44 | 4.50 |
| Gandhi | 0 | ∞ | 0.77 | 49.28 | 1.18 | 47.42 | 4.52 |
| Temple | 0 | ∞ | 0.75 | 49.39 | 1.16 | 47.47 | 4.49 |

Table III: Method 2

| Cover Image | Channel I Red | | Channel II Green | | Channel III Blue | | BPP (Bits Per Pixel) |
|---|---|---|---|---|---|---|---|
| | MSE | PSNR | MSE | PSNR | MSE | PSNR | RGB |
| Lena | 0.76 | 49.33 | 0 | ∞ | 1.17 | 47.46 | 4.50 |
| Baboon | 0.76 | 49.33 | 0 | ∞ | 1.17 | 47.45 | 4.49 |
| Gandhi | 0.80 | 49.12 | 0 | ∞ | 1.17 | 47.46 | 4.49 |
| Temple | 0.75 | 49.38 | 0 | ∞ | 1.17 | 47.46 | 4.50 |

Table IV: Method 3

| Cover Image | Channel I Red | | Channel II Green | | Channel III Blue | | BPP (Bits Per Pixel) |
|---|---|---|---|---|---|---|---|
| | MSE | PSNR | MSE | PSNR | MSE | PSNR | RGB |
| Lena | 0.51 | 51.09 | 0.64 | 50.04 | 0.78 | 49.23 | 4.49 |
| Baboon | 0.50 | 51.15 | 0.63 | 50.13 | 0.78 | 49.21 | 4.48 |
| Gandhi | 0.53 | 50.92 | 0.64 | 50.06 | 0.78 | 49.24 | 4.49 |
| Temple | 0.49 | 51.16 | 0.64 | 50.10 | 0.78 | 49.21 | 4.48 |

## A. Comparsion

Comparing the histograms of the RGB channels before and after the embedding, higher security performance was inferred. Interestingly, from different test runs we arrived at different distributions between the three channels which had no particular pattern in common. This varying pattern assured that this algorithm can be considered as pseudorandom, based on the randomness of the of the indicator channel.

The following are the stego images embedded with full embedding capacity:

Figure 2 Method I    Figure 3 Method II    Figure 4 Method III

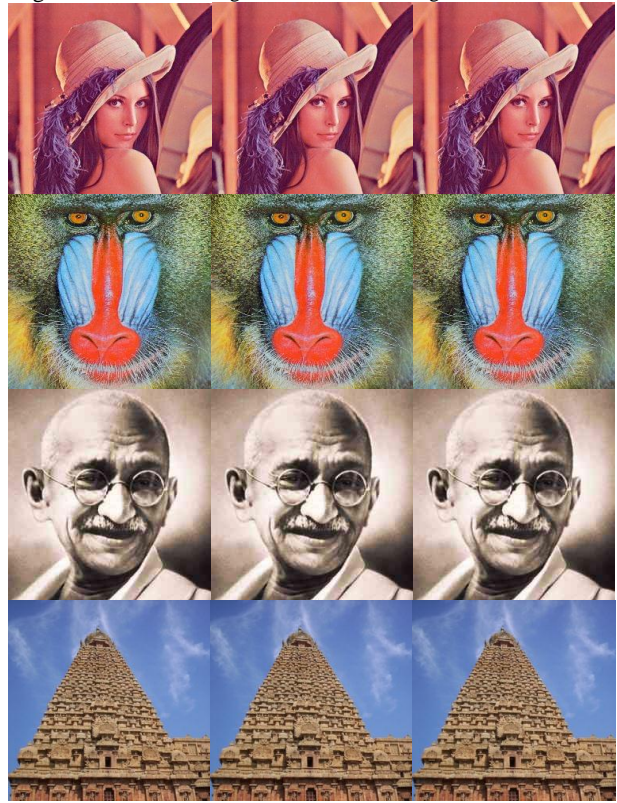

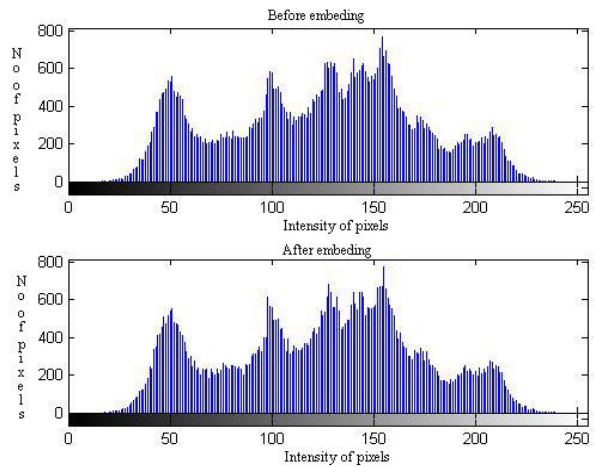

Figure 5 (a) Histogram of Lena for method -I:

Figure 5 (b) Histogram of Lena for method -II:

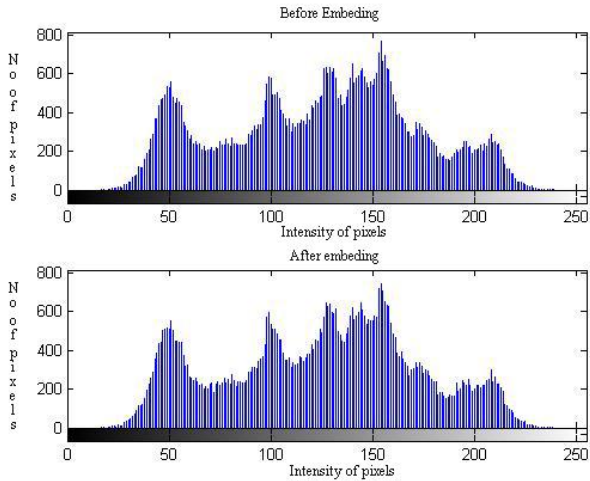

Figure 5 (a) Histogram of Lena for method - III:

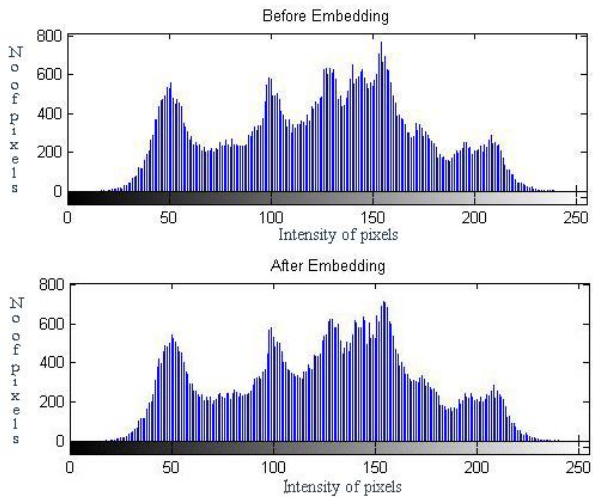

## B. Discussion

By studying the histograms of images pre and post embedding, it is inferred that:
In the first two methods, the second method gives more flexibility to the users, the indicator channel does not participate in the embedding process and hence the MSE corresponding to it is zero. This makes the image susceptible to attacks.

In the third method the MSE is uniformly distributed among all the three channels as it uses all the channels sequentially. This improves the imperceptibility and enhances the embedding capacity. The third methodology is the best among the three. By applying the OPAP the image quality is enhanced which is seen in the PSNR value.

## V. CONCLUSION

In this paper we have proposed a novel and adaptive method to embed the secret data in the cover image with high security, imperceptibility and enhanced embedding capacity. The receiver does not need the original image to extract the information. The table presented is just a prototype and can be modified by the user to any desired level, which makes the method more flexible along with improved randomness. Moreover the embedding depends completely on the nature of the pixels which is not predictable. This makes it completely adaptive and random because the nature of the pixels cannot be controlled; it is inherent of an image. Our testing and results have shown that our method III performs better when compared to the methods I and II without giving any noticeable distortions.

## ACKNOWLEDGMENT

The authors wish to thank Dr.R.Varadharajan, Professor / ECE and Dr. K.Thenmozhi Associate Dean ECE / SEEE/ SASTRA University for their valuable guidance and support.

## AUTHORS PROFILE

**R. Amirtharajan** was born in Thanjavur, Tamil Nadu province India, in 1975. He received B.E. degree in Electronics and Communication Engineering from P.S.G. College of Technology, Bharathiyar University, Coimbatore, India in 1997 and M.Tech. in Computer Science Engineering from SASTRA University Thanjavur, India in 2007. He joined SASTRA University, Thanjavur, Tamil Nadu, India (Previously Shanmugha College of Engineering) as a Lecturer in the Department of Electronics and Communication Engineering since 1997 and is now Assistant Professor, He is currently working towards his Ph.D. Degree in SASTRA University. His research interests include Image Processing, Information Hiding, Computer Communication and Network Security. So far he has published 7 Research articles in National and International journals and 1 conference paper. He has Supervised 10 Master Students and more than 100 UG projects. Currently he is working on funded project in the field of Steganography supported by DRDO, Government of India, New Delhi.

**Sandeep Kumar Behera, Motamarri Abhilash Swarup and Mohamed Ashfaaq K** are final year B.Tech. students in the Department of Electronics and Communication Engineering, School of Electrical & Electronics Engineering, SASTRA University. Apart from excellent Academic record, they studied an open elective course in Information Hiding. So far these students presented around 10 paper and project presentations in various National Level Student Symposiums and Project Presentation Contests and 2 National Conference in colleges and in Universities in India.

**Dr. John Bosco Balaguru Rayappan** was born in Trichy, Tamil Nadu province, India in 1974. He received the B.Sc., M.Sc. and M.Phil. Degree in Physics from St. Joseph College, Bharathidasan University, Trichy and Ph.D. in Physics from Bharathidasan University, Trichy, Tamil Nadu India in 1994, 1996, 1998 and 2003, respectively. He joined the faculty of SASTRA University, Thanjavur, India in Dec 2003 and is now working as Professor in School of Electrical and Electronics Engineering at SASTRA University, Thanjavur, Tamil Nadu, India. His research interests include Lattice Dynamics, Nanosensors, Embedded System and Steganography. So far he has published 20 Research articles in National and International journals and 14 conference papers. He has Supervised 25 Master Students and Supervising 3 Ph.D. Scholars. Currently he is working on four funded projects in the fields of Nanosensors and Steganography supported by DST and DRDO, Government of India, New Delhi.